\documentclass[12pt]{article}
\usepackage[left=2cm,top=2.5cm,right=2cm,bottom=3cm,nohead]{geometry}
\usepackage{amsmath,amsfonts,amsthm,amssymb,times, bm}
\input amssym.def
\input amssym.tex
\def\ben{\begin{equation}}
\def\een{\end{equation}}

\def\bea{\begin{eqnarray}}
\def\eea{\end{eqnarray}}

\def \p{\partial} 
\usepackage{epsfig}

\begin{document}

\hfuzz=100pt
\title{Circular, stationary profiles emerging in unidirectional abrasion}
\author{G. Domokos
\footnote{Department of Mechanics, Materials, and Structures,
Budapest University of Technology and Economics,
M\"ugyetem rkp.3,Budapest 1111, Hungary},\,
G.~ W.~ Gibbons \footnote{D.~A.~M.~T.~P., Cambridge University,
Wilberforce Road, Cambridge CB3 0WA, U.K.}
\, and A.A. Sipos$^{*}$}

 \maketitle
\begin{abstract}
We describe a PDE model of bedrock abrasion by impact of moving particles and
show that by assuming unidirectional impacts the modification of a geometrical PDE due to Bloore \cite{Bloore}
exhibits circular arcs as solitary profiles. We demonstrate the existence
and stability of these stationary, travelling shapes by numerical experiments 
based on finite difference approximations. Our simulations show that ,depending on initial
profile shape and other parameters, these circular profiles may evolve via long
transients which, in a geological setting, may appear as non-circular stationary profiles.
\end{abstract}


\section{Introduction}  

\subsection{Motivation and references}
Bedrock abrasion by impact of moving particles is a dominant process shaping obstacles in river channels \cite{Sklar1}. 
Abrasion is mostly due to saltating particles, a broad description of their mechanistic properties is presented in \cite{Sklar2}. 
Our goal is to identify and predict the geometry of obstacle profiles created by this process. In a recent paper, a discrete random model 
describing bedrock profile abrasion was developed
and its predictions compared with experiment \cite{Sipos}.  
The main observation was that \em stationary profiles \rm emerged
both in the laboratory experiments and the matching discrete
simulations. This phenomenon has also been observed in Nature \cite{Wilson},\cite{WilsonHovius},\cite{Wilson2}.
The aim of the present paper is to present a simple partial differential
equation capturing the essence of the physical process
and investigate whether  it admits stable travelling front solutions
which are compatible with the numerical and experimental results obtained
in \cite{Sipos} and observed in \cite{Wilson},\cite{WilsonHovius}.

\subsection{Basic assumptions and notations}
If $(x,y,z)$ are Cartesian coordinates we take 
$y$ along the vertical direction. We assume, for simplicity, that 
the system is uniform in the horizontal direction  perpendicular 
the direction of motion of the abrading particles. Thus the     
bedrock  profile starts off and remains independent of the $z$ coordinate
and is therefore  given by a single function   
$y=y(x,t)$  of space and time. The tangent to the profile makes an angle
$\psi$ with respect to the horizontal, so the inward normal to the profile
makes an angle $\psi-\frac{\pi}{2}$ with the horizontal. We also introduce
\ben \label{notation}
y'=\p y/\p x=\tan \psi
\een  
and assume, following \cite{Sipos}, that 
\begin{itemize}
\item small abraders are incident  from the left making an angle 
$\phi$ with the horizontal.
\item the angle of the colliding face of large abraders
is uniformly distributed between $0$  and $\frac{\pi}{2}$,  i.e.
the non-uniform flight of abraders does not change this term.
\end{itemize} 
\begin{figure}
\begin{center}
\includegraphics[width=120 mm]{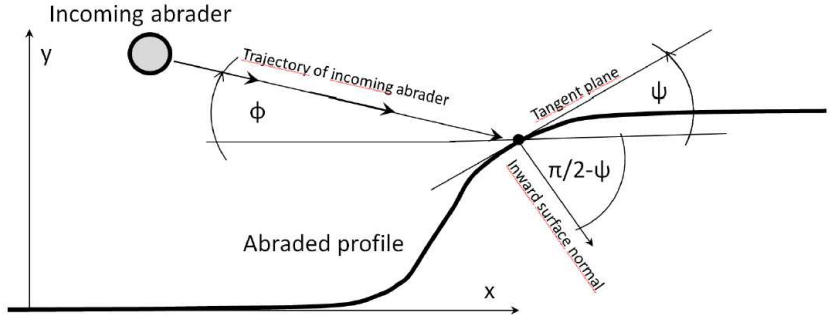}
\end{center}
\caption{Basic notations}\label{Fig1}
\end{figure}
As time proceeds, a point on the profile with
constant $x$ coordinate moves 
with vertical velocity  given by 
\ben
v_{\rm vertical}= \frac{\p y}{\p t}= \dot y,
\een
the partial derivative being taken at fixed $x$. The velocity 
of the same point along the inward normal is:
\ben
v_n=-\dot y \cos \psi.
\label{one}
\een
Abrasion due to collisions with incoming abraders with isotropic, uniform distribution has been first derived by Bloore \cite{Bloore}, for the detailed
derivation of the planar case see also \cite{DomokosSiposVarkonyi2}.
Under the assumption of unidirectional abraders, Bloore's PDE can be readily modified by the inclination factor to obtain
\ben \label{bloore}
v_n= (v+ b\rho) \sin (\psi -\phi), 
\een
where $\rho=-y''(1+y'^2)^{-3/2}$ is the geometric curvature
and $v, b$ are positive constants.
Using the latter expression as well as (\ref{one}), we obtain 
\ben \label{four}
 -\dot y\cos \psi=(v+ b\rho) \sin (\psi -\phi)=(v+ b\rho)(\sin \psi \cos \phi - \cos \psi \sin \phi)
\een 
Using (\ref{notation}), from (\ref{four}) we obtain
\ben
-\dot y=(v+ b\rho) (\cos \phi y'-\sin\phi).
\een
Without restricting generality, we assume $\phi=0$ (i.e. that the $x$ axis is aligned with the flight
direction of the abraders) to obtain
\ben
-\dot y=y'(v+b\rho)=y'(v+by''(1+y'^2)^{-\frac{3}{2}}).
\label{basiceqn}
\een
Note that $L=b/v$ is the only independent length scale of the equation. 
Rather than descibe the profile in terms of $y$ considered as a function of
$x$ and $t$ we may regard $x$ as a  function of $y$ and $t$: $x=x(y,t)$, so we obtain: 
\ben
\dot x =(v+b\rho)=(v+bx''(1+x'^2)^{-\frac{3}{2}})\,
\label{basiceqn2}
\een
where now $\dot x= \p_t x(y,t)$ keeping $y$ fixed and $x^\prime = \p_y (y,t),$ keeping $t$ fixed.

Our equation is a model of collision-based abrasion,
and the first (constant) term corresponds to the abrasion caused by small abrading particles whereas the second (curvature)
term is significant if the abrader is large (cf. \cite{Bloore},\cite{Sipos},\cite{DomokosSiposVarkonyi2}). Based on this
physical background, in forthcoming numerical simulations we will assume that on non-convex parts of the profile the curvature term vanishes (as large abraders are not
likely to hit concave domains). Accordingly, the modified equation reads:
\ben \label{nonconvex}
-\dot y=y'(v+b\bar \rho), \mbox{ where} \quad \bar \rho =\rho \mbox{ if } \rho>0 \quad \mbox{and} \quad \bar \rho =0 \mbox{ if }\rho \leq 0.
\een

\section{Existence of travelling waves}\label{travelling} 
Note that (\ref{basiceqn}) is first order in time and second
order in space and translationally invariant in  time. It is also
invariant under translations in space, both vertical and horizontal.
Being first order in time (\ref{basiceqn}) defines a ``flow'' on 
the infinite dimensional space of profiles  
such that if we are given a smooth  initial profile $y(x,0)= y_{\rm initial}(x)$
say, then, at least  for some finite time $y(x,t)$ is 
uniquely determined. In what follows we shall investigate 
whether,  for  smooth  initial  Cauchy  data 
such that $y_{\rm initial}(x)$  is monotonic increasing and 
tends rapidly to constant values for large negative and positive
values of $x$, that   the solution $y(x,t)$ ultimately tends
to an exact  travelling front solution. 
If it does, then we shall derive the form it must take.
If we make the common travelling wave {\it ansatz}
\ben \label{ansatz}
y=f(x-ct) 
\een 
and after substituting into (\ref{basiceqn}) solve the resulting ordinary differential equation for $f(s)$ 
we find that  the solutions are all of the form of portions 
of circles or straight lines which move to the right
with speed
\ben \label{speed}
c=v+\frac{b}{a}
\een
where $a$ is the radius of the circle. Two special cases
should be mentioned:
For straight lines of course we have $a=\infty$ and $c=v$.
In case of $b=0$ we have
\ben
\frac{\p y}{\p t}  = - v \frac{\p y}{\p x}
\een 
whose general solution is 
\ben
y=f(x-ct) \,,\qquad {\rm with} \qquad c=v  
\een
thus if $b=0$ then \em any initial shape \rm moves with to the right with speed $c=v$.
These (local) travelling wave solutions travelling with speed given in (\ref{speed}) may easily be derived from the alternative 
form of the governing equation (\ref{basiceqn2}) 
by making the ansatz $x(y,t)= ct + x(y).$

The local solutions, made up of circles
and straight lines may be patched together to
give global solutions satisfying the boundary conditions.  
Matching conditions should be observed carefully. E.g. only
segments with identical curvature may be patched together, otherwise
the propagation speed $c$ will suffer a jump.
If we use boundary conditions representing  a 'cliff´
(cf Figure \ref{Fig1}) then we have two options:

\begin{itemize}

\item (a)  We satisfy horizontal tangency as boundary condition at both lower and upper end.
In this case if we apply (\ref{nonconvex}) prescribing that the curvature term vanishes on concave parts.
In this case there is no exact travelling wave solutions as the two ends of the cliff can not be
joined by a curve with constant curvature. The only exact solutions
are half circles,  however these contours
are not  a feasible representation of real cliffs.

\item(b) We satisfy the horizontal tangency only at the upper end. At the lower
end we do not prescribe a boundary condition, i.e. we essentially cut off
the profile at $y=0$.  This admits a one-parameter ($R$) family circular arcs
which would be exact travelling solutions. This approach could be justified by
considering that a vertex on the upper end is convex (thus it will be abraded
by large abraders) whereas a vertex on the lower end is concave so large
abraders to not smear it out. Small abraders, corresponding to the constant
(Eikonal ) term do not abrade any vertices's. So, admitting a concave vertex at
the lower end does not seem to contradict the physical assumptions.
This type of boundary condition has been implemented numerically.
Figure \ref{Fig2} shows the evolution
of a circular arc under (\ref{basiceqn}) and boundary condition (b). The numerical simulation was
based on a standard finite difference scheme and we can observe that
the circular arc remains invariant and moves by uniform translation.
\begin{figure}
\begin{center}
\includegraphics[width=120 mm]{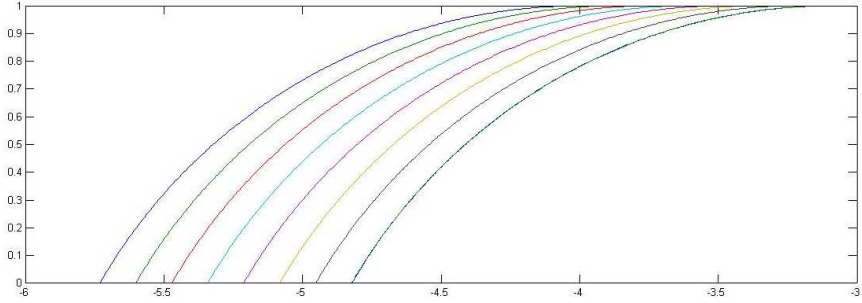}
\end{center}
\caption{Circular arc with radius $a =2$ and horizontal tangent at $y=1$ evolving under (\ref{basiceqn}) with $L=b/v=0.6$. Numerical simulation by standard finite difference scheme,
using $N_s=300$ spatial subdivisions and $\Delta t=10^{-5}$ timestep, profiles plotted after $N_t=10^4$ timesteps each. Observe that the profile moves with uniform translation, circular
arc remains invariant under (\ref{basiceqn}). }\label{Fig2}
\end{figure}

\end{itemize}

\section{Stability}
Next we investigate the stability of travelling waves. After finding analytical evidence in subsection \ref{linear} for their local stability we
run numerical simulation in subsection \ref{numerical} to test their global attractivity.

\subsection{Linear stability: analytical results} \label{linear}

Suppose $y(x,t)$ is a solution of 
(\ref{basiceqn}) when we assume the profile is convex, i.e. $\frac{\p ^2 y}{\p x^2}
$ is negative, so that  
\ben 
 \frac{\p y}{\p t}  = 
-(
v-b \frac{ \frac{\p ^2 y}{\p x^2}  }
{
\Bigl(1+ (\frac{\p y}{\p x}) ^2 \bigr ) ^{\frac{3}{2} }
 }
) \frac{\p y}{\p x} \label{eqn}    
\een
with $b>0$. We consider a nearby solution $y(x,t)+ \epsilon(x,t)$ 
and substitute into  (\ref{eqn}) and expand to lowest order in $\epsilon$
we get
\ben
 \frac{\p \epsilon}{\p t}  = 
- \Bigl \{ 
v-b  \frac{ \frac{\p ^2 y}{\p x^2} (1-2 (\frac{\p y}{\p x})^2   }
{\bigl(1+ (\frac{\p y}{\p x}) ^2 \bigr ) ^{\frac{5}{2} } }    \Bigr \} 
 \frac{\p \epsilon}{\p x} +b \frac{\p y}{\p x} \frac{1}
{\bigl(1+ (\frac{\p y}{\p x}) ^2 \bigr ) ^{\frac{3}{2} }} \frac{\p ^2  \epsilon }{\p x^2}
\een
This equation is of the form
\ben \label{perturbation}
 \frac{\p \epsilon}{\p t}  +u(x,t) \frac{\p \epsilon}{\p x}
= \kappa(x,t)  \frac{\p ^2  \epsilon }{\p x^2}
\een
with convection velocity 
\ben \label{convection}
u(x,t) = 
v-b  \frac{ \frac{\p ^2 y}{\p x^2} (1-2 (\frac{\p y}{\p x})^2   }
{\bigl(1+ (\frac{\p y}{\p x}) ^2 \bigr ) ^{\frac{5}{2} } }  
\een
and diffusivity 
\ben \label{diffusion}
\kappa(x,t) = b \frac{\p y}{\p x} \frac{1}
{\bigl(1+ (\frac{\p y}{\p x}) ^2 \bigr ) ^{\frac{3}{2} }}.
\een
Essentially, the first term convects the disturbances with a positive space
and time dependent  speed $u(x,t)$ and the second term is a diffusion term
which causes the disturbance to spread out and decrease in amplitude.
This is independent  of the particular profile $y(x,t)$. In case of straight lines
\ben
y=A(x-vt) = As, 
\een
based on (\ref{convection}) and (\ref{diffusion}) we find the convection velocity and diffusivity
\ben
u(x,t)=v, \qquad \kappa(x,t)=b \frac{A}{1+A^2}.
\een
We can aslo rewrite (\ref{perturbation}) using 
or using the co-moving variables $s,t$
\ben
\frac{\p \epsilon}{\p t} =  b \frac{A}{1+A^2} \frac{\p ^2  \epsilon }{\p s^2} 
\een
which is the standard diffusion equation. So we can conclude that all travelling profiles are locally stable, moreover, for
straight lines we have explicitly determined the diffusion and convection terms.

\subsection{Nonlinear stability: numerical results} \label{numerical}

\begin{figure}
\begin{center}
\includegraphics[width=120 mm]{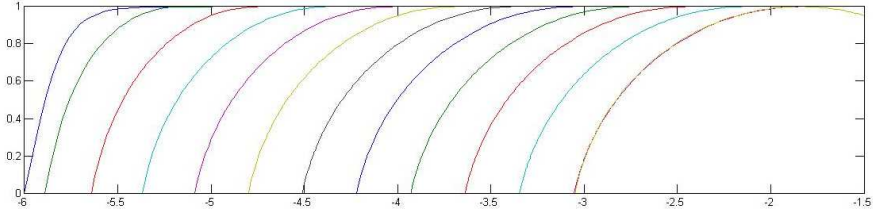}
\end{center}
\caption{Initial profile given in (\ref{halftanh}) with $\lambda=5$  evolving under (\ref{basiceqn}) with $L=b/v=0.6$. Numerical simulation by standard finite difference scheme,
using $N_s=135$ spatial subdivisions and $\Delta t=10^{-5}$ timestep, profiles plotted after $N_t=2*10^4$ timesteps each (Since the shape changes faster initially, 
between the first an second profile an intermediate stage
at 10.000 steps was inserted). Observe that profile approaches circular arc, next to
last profile exact circle is plotted for better comparison. }\label{Fig3}
\end{figure}

Circular profiles are of key interest from the
point of view of geological applications, we investigate their stability numerically.
Figure \ref{Fig2} illustrated the \em local \rm stability of circular
arcs: by taking such an initial condition the circualr shape
was preserved by the PDE. In te next step we consider the convex profile
\ben \label{halftanh}
y_{\rm initial}(x)=\tanh(\lambda x), \quad x\geq 0
\een
which satisfy $y(0)=0,\quad y(\infty)=1$ and we evolve these
profile numerically considering the boundary condition
(b) described in Section \ref{travelling}. The results
are shown in Figure \ref{Fig3} and we can observe how the
initial profile approaches a circular arc. For better
comparison the last profile is plotted next to an exact circle.

\section{Comparison to experimental data}
\begin{figure}
\begin{center}
\includegraphics[width=100 mm]{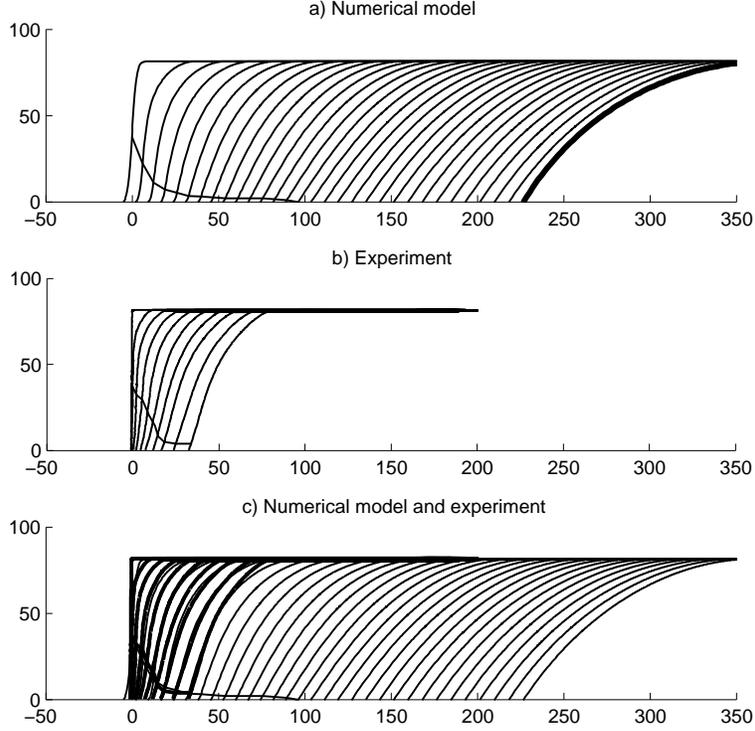}
\end{center}\caption{Comparison of experimental and numerical data. Experimental data is based entirely on (\cite{Wilson}).
Initial profile given in (\ref{expt_num})  evolving under (\ref{basiceqn}) with $L=b/v=2.0$. Numerical simulation by standard finite difference scheme,
using $N_s=750$ spatial subdivisions and $\Delta t=10^{-6}$ timestep, profiles plotted after $N_t=12500$ timesteps each. Observe that numerical and
experimental profile show very similar evolution. Profile approaches circular steady-state circular geometry with radius $a=131.5$ only much beyond
the range of the physical experiment. 
Location of inflection point is plotted both on experimental and numerical profiles. }\label{Fig4}
\end{figure}

Here we rely on laboratory experimental data
obtained by Wilson \cite{Wilson} \cite{Lave} using an annular, recirculating flume \cite{Lave}. In an experiment with this flume \cite{Wilson}, single cuboid marble blocks, 10,0 cm tall and 20 cm long, and spanning the 26 cm width of the flume, were fixed on the channel base as a flow obstruction. In our model this would correspond to
a Heaviside step function as initial profile shape. The flume was loaded with sorted limestone pebbles to act as bedload abrasion tools, while water discharge was held constant to produce a flow speed of ~3ms-1.  Under these flow conditions, the limestone pebbles moved by rolling and low saltations in a layer of one to two grain diameters depth along the flume base. Particles moved up the stoss (upstream facing) surface of an obstacle in one or several short hops, and launched off this face clearing the remainder of the obstacle. Resultant erosion of bedrock obstacles was measured using three dimensional laser scanning at intervals throughout experiment runs that lasted 400-690 minutes. Abrasion of obstacles was dominantly on the stoss surface whose initially square cross section evolved to an unpstream facing, almost convex surface. The lower 18.5 mm of the stoss surface were prone to edge chipping, and have been eliminated from consideration, the plotted profile is 81.5mm high. Rates of horizontal erosion were peaking near the top of the obstruction, and decreasing systematically towards the base, resulting in progressive reclining and convex-up rounding of the stoss side. Over time horizontal erosion converged to a common rate at all heights above the flume base,  thus, the obstacle stoss faces achieved a time-independent form that advanced downstream into the obstacle. These attributes are shared with bedrock obstacles in the Liwu River, Taiwan, that have been studied extensively \cite{Hartshorn}. We have set up numerical simulation applying our model to match the flume experiment. We approximated the initial profile by 
\ben \label{expt_num}
y=41.5\tanh(0.4x)+40
\een
The evolution of (\ref{expt_num}) under (\ref{basiceqn}) is plotted in Figure \ref{Fig4}/a. Since this is a non-convex profile, 
we used the modified abrasion law defined in (\ref{nonconvex}). We can observe that in this case the circular profiles are reached only after
long transients, i.e. the inflection point (trajectory plotted in Figure \ref{Fig4}/a) lingers for an extended period of time above the $x$ axis. 
This may account for the geologial observation of non-convex 'steady-state´ profiles
which correspond actually to the long transients before the circular shape settles in. In Figure \ref{Fig4}/b experiments are shown and Figure \ref{Fig4}/c.

\section{Conclusions}
In this paper we set up the simple PDE model (\ref{basiceqn}) for abrasion under unidirectional (parallel) impacts, based on Bloore's model \cite{Bloore} for isotropic abrasion. By using the travelling wave
ansatz (\ref{ansatz}) we found circular travelling solutions. After establishing their local (linear) stability analytically, we tested their global attractivity by finite difference simulations
and found that they appear to be globally stable. We compared our numerical results to laboratory experiments of Wilson \cite{Wilson} and found good agreement. Our simulation showed
that nonconvex profiles may approach their final, circular steady-state geometry via long transients as inflection points may linger above the $x$ axis for extended periods of time.
This fact may account for the geological observation of (almost) stationary, non-convex profiles in riverbeds.

\section{Acknowledgments}
This research was supported by the Hungarian National Science Foundation Grant OTKA K104601.

\end{document}